\documentclass{article}
\usepackage[preprint]{spconf}
\usepackage{amsmath,amsfonts,amssymb,amsthm,graphicx}
\usepackage{url}
\usepackage{hyperref}
\usepackage{enumitem}

\makeatletter
\def\blfootnote{\gdef\@thefnmark{}\@footnotetext}
\makeatother

\usepackage{balance}
\usepackage{bm}

\newcommand{\RR}{\mathbb{R}}

\usepackage{cite}
\usepackage{xcolor}
\usepackage{multirow}
\usepackage{etoolbox,siunitx}
\robustify\bfseries
\usepackage{booktabs}

\usepackage{caption}
\usepackage{subcaption}

\let\OLDthebibliography\thebibliography
\renewcommand\thebibliography[1]{
  \OLDthebibliography{#1}
  \setlength{\parskip}{.5pt}
  \setlength{\itemsep}{.5pt plus 0.3ex}
}

\title{A fully differentiable model for \\unsupervised singing voice separation }
\name{Gaël Richard \quad Pierre Chouteau \quad Bernardo Torres 
}
\address{LTCI, T\'el\'ecom Paris, Institut polytechnique de Paris, France}

\copyrightnotice{\copyright 2024 IEEE. To appear in {\it Proc.\ ICASSP2024, } Personal use of this material is permitted. Permission from IEEE must be obtained for all other uses.}

\begin{document}
\ninept
\maketitle
%
%
\begin{abstract}

A novel model was recently proposed by Schulze-Forster et al. in \cite{schulze-forster_unsupervised_2023} 
for unsupervised music source separation. This model allows to tackle some of the major shortcomings of existing source separation frameworks. Specifically, it eliminates the need for isolated sources during training, performs efficiently with limited data, and can handle homogeneous sources (such as singing voice). But, this model relies on an external multipitch estimator and incorporates an Ad hoc voice assignment procedure. In this paper, we propose to extend this framework and to build a fully differentiable model by integrating a multipitch estimator and a novel differentiable assignment module within the core model. We show the merits of our approach through a set of experiments, and we highlight in particular its potential for processing diverse and unseen data. 


\end{abstract}

\section{Introduction} 
\blfootnote{This work was funded by the European Union (ERC, HI-Audio, 101052978). Views and opinions expressed are however those of the author(s) only and do not necessarily reflect those of the European Union or the European Research Council. Neither the European Union nor the granting authority can be held responsible for them.}


Music Source Separation (MSS) \cite{canoMusicalSourceSeparation2018}, the task of estimating the individual music signals when only a mixture is available,  has become an indispensable tool in various applications, ranging from remixing tracks and audio transcription to music recommendation and melody extraction for beginner musicians. 
The field's state-of-the-art now relies on leveraging deep neural networks (DNNs) for this task \cite{defossezDemucsDeepExtractor2019, stoterOpenunmixaReferenceImplementation2019, takahashiD3netDenselyConnected2020}. DNNs offer the advantage of working directly on raw audio signals, bypassing the need for hand-crafted features. However, these methods have their own shortcomings.
 Firstly, they mainly rely on supervised training, meaning that isolated sources must be accessible and available for training. Secondly, these models are often excessively complex, requiring large amounts of data and computational resources. Finally, although these models perform well on specific datasets, they can run into difficulties when the input data is homogeneous, such as when dealing with sets of choruses of similar voices.

Given these limitations, methods that can operate effectively without access to isolated sources have emerged. Techniques like Non-negative Matrix Factorization (NMF) \cite{leeLearningPartsObjects1999} have shown promise, but often rely on prior information \cite{ewertUsingScoreinformedConstraints2012}  or heavy assumptions about the sources\cite{durrieuMusicallyMotivatedMidlevel2011}, and might have low flexibility due to a pre-defined number of spectral templates. Unsupervised deep-learning-based approaches are promising but only few works address homogenous or correlated sources. 

This paper builds upon the work of Schulze-Forster et al. \cite{schulze-forster_unsupervised_2023}, which proposes an unsupervised DNN model that has shown state-of-the-art performance in separating choral singing. 
We expand their work by integrating the multi-F0 estimation and voice assignment modules as differentiable blocks within the model and by proposing a novel method for differentiable F0 contour extraction. We then obtain an end-to-end, fully differentiable model for unsupervised source separation. We also conduct an extensive experimental validation to demonstrate the efficacy of our methods.




The paper is organised as follows: we recall our baseline unsupervised source separation method in Section \ref{sec:unsupervised} before presenting in Section \ref{sec:differentiable} our new fully differentiable approach. The experiments and results are respectively presented in Sections \ref{Sec:Experiments}  and \ref{Sec:Results}. Finally, we suggest some conclusions in Section \ref{sec:conclusion}.

\section{Unsupervised music source separation}
\label{sec:unsupervised}

The original model proposed in \cite{schulze-forster_unsupervised_2023}, referred to as UMSS, was shown to be efficient for complex source separation problems in which individual sources are not available for training,  the sources are homogeneous (only singing voices), or only a limited amount of mixture recording data is obtainable. The approach is inspired by the recent hybrid deep learning paradigm, which integrates signal processing models in DNNs to incorporate domain knowledge \cite{shlezinger2023model,engel_ddsp_2020}. In UMSS, each source is represented with a differentiable parametric source-filter model. During training, the task of the DNN is to re-synthesize the observed mixture as a sum of the sources by estimating the source parameters given their fundamental frequencies. The source estimates can be obtained either directly as the synthesized sources or by filtering the initial mixture with soft masks obtained from the synthesized source signals. The latter strategy obtains the best overall results and is then selected in this work.   

\begin{figure}[!htbp]
    \centering
    \includegraphics[width=0.98\linewidth]{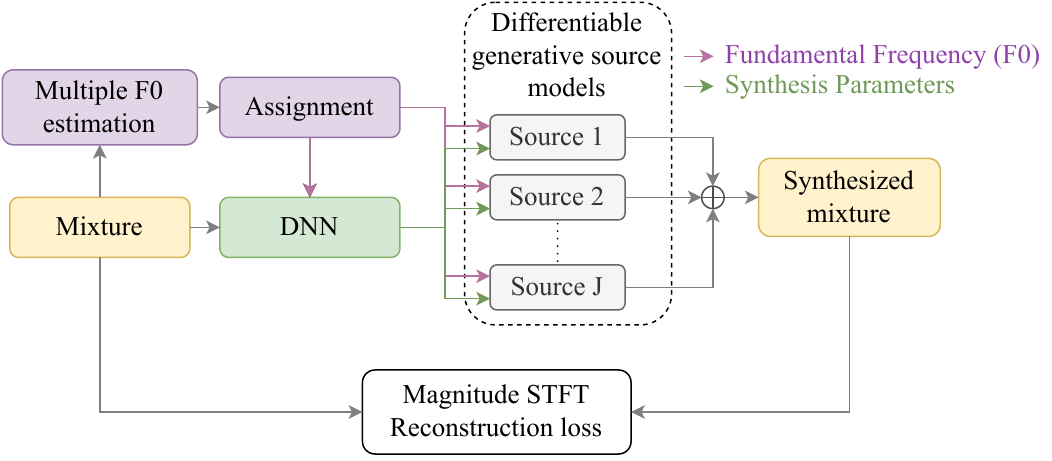}
    \caption{Overview of the unsupervised music source separation approach proposed in \cite{schulze-forster_unsupervised_2023}.
    }
    \label{fig:première_implémentation}
\end{figure}


The model used in \cite{schulze-forster_unsupervised_2023} to obtain source fundamental frequencies was given in \cite{cuesta_multiple_2020} and performs multi-F0 extraction by first processing a spectral representation through a DNN, and then converting the output multi-frequency salience map to F0 contours. In \cite{schulze-forster_unsupervised_2023}, a voice assignment heuristic based on temporal pitch continuity was further added to enable training of the source separation model. A closer look at this pipeline reveals that both salience-map-to-F0 and voice assignment operations are not differentiable. 
Addressing these issues, we introduce differentiable extensions to these non-differentiable blocks, achieving a fully differentiable architecture.

\section{A fully differentiable model}
 \label{sec:differentiable}
 



\subsection{The complete model}
\begin{figure}[!htpb]
  \centering
  \includegraphics[width=0.95\linewidth]{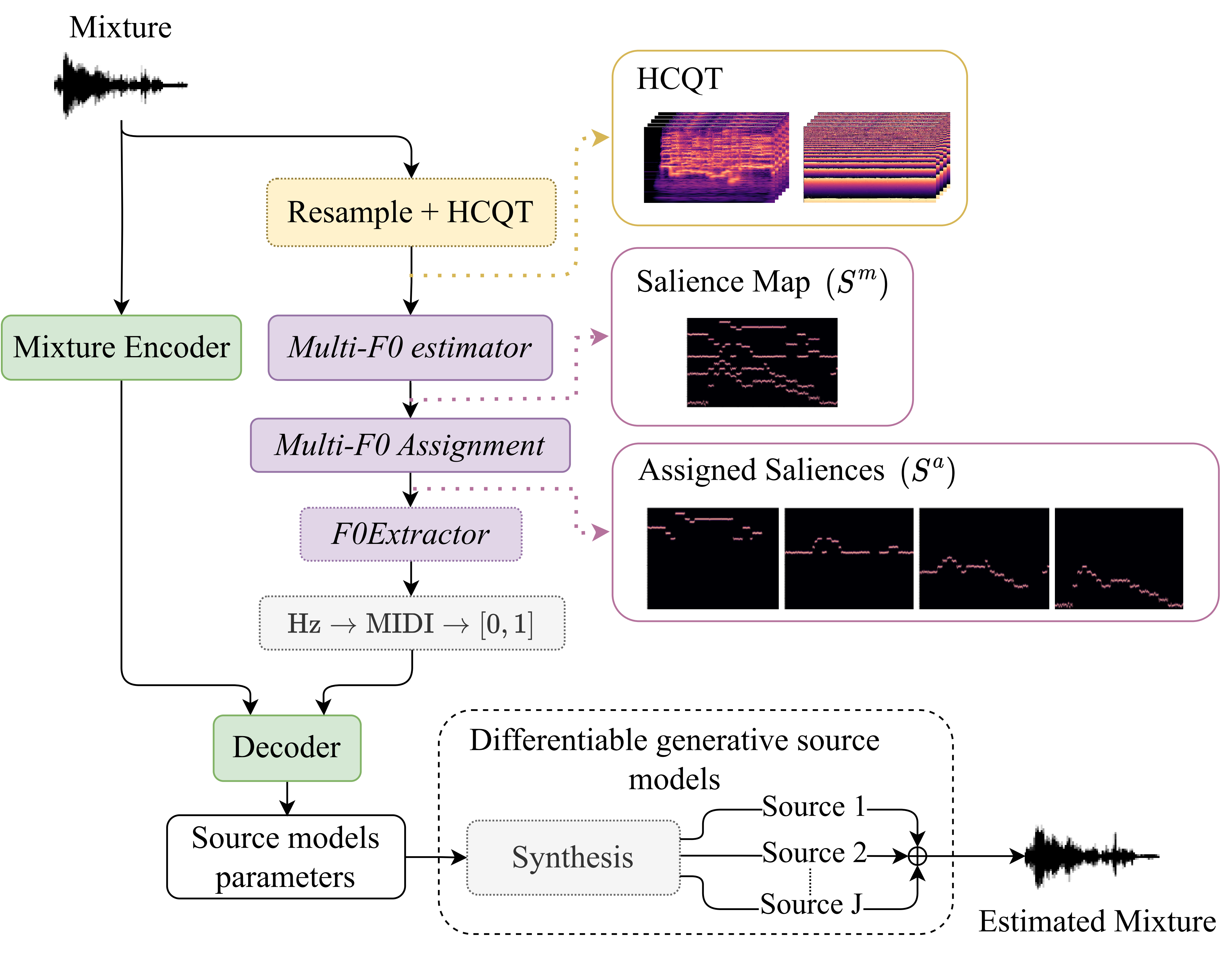}
  \caption{
  Fully differentiable architecture for joint learning. 
  The blocks in purple are those that have been added to the initial model.
  } 
  \label{fig:ArchitectureComplete}
\end{figure}

The complete model is shown in Fig. \ref{fig:ArchitectureComplete}. It is based on the unsupervised model described above \cite{schulze-forster_unsupervised_2023} but with the integration of the multi-F0 estimation and vocal assignment as three differentiable blocks (displayed in purple on Fig. \ref{fig:ArchitectureComplete}). The resulting architecture is then fully differentiable and can be trained end-to-end. 

More precisely, the proposed architecture takes as input a 4-second audio mixture which is processed in parallel in two branches: 1) the first branch is based on the encoder of \cite{schulze-forster_unsupervised_2023}, which extracts the main characteristics of the observed audio. 2) The second branch is dedicated to the estimation of multiple fundamental frequencies. It consists of three blocks: the first estimates a multi-frequency salience map from the observed audio (\textit{Multi-F0 estimator} from \cite{cuesta_multiple_2020}); then, this multi-frequency salience map is converted to  assigned salience maps, corresponding to the time-frequency representation of the fundamental frequencies of each voice (\textit{Multi-F0 Assignment}, from \cite{cuesta_voice_2022}); finally, our new contour extraction method is applied to each assigned salience map to obtain the F0 sequences of each source over time (\textit{F0Extractor}).




The full loss used to optimize the training is given by:
\begin{equation}
    \mathcal{L_{\text{full}}} =
    \mathcal{L_{\text{rec}}} + \alpha \mathcal{L}_1 +   \beta\mathcal{L}_2 + \gamma \mathcal{L}_3,
    \label{eq:full_loss}
\end{equation}

\noindent where 
\textit{$\mathcal{L}_{\text{rec}}$} is the commonly adopted Multi-Scale Spectral Loss \cite{engel_ddsp_2020} computed between the magnitude STFT's of the input mixture $x$ and estimated mixture $\hat x$ (the sum of the estimated sources):

\begin{equation}
    \mathcal{L}_{\text{rec}}(x, \hat x)=\sum_{n \in \mathcal{N}} \mathcal{L}_{\text{lin}}^n + \mathcal{L}_{\text{log}}^n,
\end{equation}\label{eq:msstft}

\noindent with $\mathcal{L}_{\text{lin}}^n = \left\|\operatorname{STFT}_{n}(x) - \operatorname{STFT}_{n}(\hat x)\right\|_{1}$ for scale (window size) $n$ and $\mathcal{L}_{\text{log}}^n =  \left(\left\|\log (\operatorname{STFT}_{n}(x))- \log(\operatorname{STFT}_{n}(\hat x))\right\|_{1}\right)$. 

$\mathcal{L}_1$ is a loss between the sum of the individual assigned salience maps $S_j^{a}$ $\in \RR^{L \times M}$ and the unassigned multi-frequency salience map $S^{m}$ $\in \RR^{L \times M}$, to constrain 
the sum of assignments to equal the input multi-frequency salience map $S^{m}$: 
\begin{equation}
    \mathcal{L}_1 = \operatorname{MSE}(\sum^{J}_{j=1} S_j^{a}, S^{m}),
\end{equation}\label{eq:l1}

\noindent where MSE denotes the \textit{Mean Squared Error}.

$\mathcal{L}_{2}$ is used to force the assignment model to return frequencies only within a predefined range for each voice (Soprano [$260$Hz-$880$Hz]; Alto [$190$Hz-$660$Hz]; Tenor [$145$Hz-$440$Hz]; Bass [$90$Hz-$290$Hz]) \cite{scirea_evolving_2015}. This is achieved by multiplying each assigned salience map $S_j^{a}$ by a mask corresponding to the desired interval and computing
the MSE between the two salience maps, before and after masking. 
\begin{equation}
    \mathcal{L}_2 = \sum^{J}_{j=1} \operatorname{MSE}(S_j^{a}, S^{\text{mask}}_j)
\end{equation}\label{eq:l2}

$\mathcal{L}_{3}$ is used to force the assigned salience maps to return only one voice, by means of the MSE between the assigned salience maps $S_j^{a}$ and their binary reconstructions $S_j^{b}$.
\begin{equation}
    \mathcal{L}_3 = \sum^{J}_{j=1} \operatorname{MSE}(S_j^{a}, S_j^{b})
\end{equation}\label{eq:l3}



\vspace{-0.4cm}

\subsection{Differentiable voice assignment}

\begin{figure}[]
  \centering
  \includegraphics[width=0.9\linewidth]{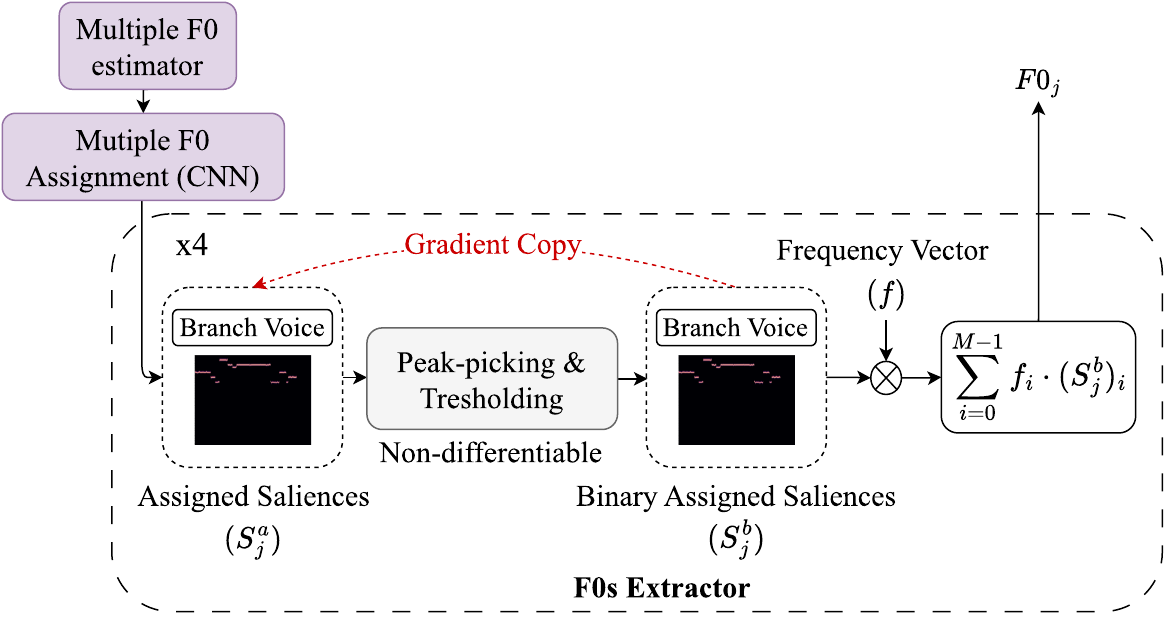}
  \caption{Extraction of F0 sequences from assigned salience maps. The gradient is copied from $S^b$ to $S^{a}$ during the \textit{backward} pass. The weighted sum of the binary salience maps is performed on the frequency axis, resulting in frame-level F0 estimations.} 

  \label{fig:rec_explanation}
\end{figure}

We describe herein the new differentiable voice assignment process in more detail.
As depicted in Fig. \ref{fig:rec_explanation}, the overall extraction of the sequence of F0s values for each source includes a number of steps. Starting from an estimated multi-frequency salience map, the \textit{multi-F0 Assignment} module assigns it to each voice in the form of an assigned salience map.  For this purpose, we exploit the model proposed by \cite{cuesta_voice_2022} which is, to the best of our knowledge, the only DNN-based model suitable for multiple singing voices. 
Then, it is necessary to convert these assigned salience maps to sequences of F0 values  for each voice. This is classically done by peak-picking followed by thresholding, which are non-differentiable operations.


To cope with this problem, we follow a similar strategy than used in 
VQ-VAE for gradient propagation of a non-differentiable function \cite{van_den_oord_neural_2017}.  First, from the assigned salience maps $S^{a}$, we reconstruct proxy binary salience maps $S^b = \{S^b_j\}$, where the binary map for source $j$,  $ S^b_j\in \RR^{L \times M}$, has the same dimension as the corresponding $S^{a}_i$. $S^b$ contains ones in the frequency bin selected after peak-picking and zeros on all other bins. The F0 contour is then computed as $F0_j = \sum^{M-1}_{i=0} f_i \cdot (S_{j}^{b})_i $, where $f$ is a frequency vector containing the corresponding bin frequencies in Hz. During the \textit{forward} process, the estimation of the sources proceeds normally. In the \textit{backward} pass, however, gradients from $S^b$ are copied to $S^{a}$. 

\subsection{Implementation details and training parameters}

The whole architecture is designed for signals sampled at $16$ kHz except for the multi-F0 estimation model \cite{cuesta_multiple_2020} which works at $22$~kHz. To cope with this, the input to the multi-F0 estimation module is upsampled to $22$ kHz.
Harmonic Constant-Q transform (HCQT) \cite{bittner_deep_2017} is computed on-the-fly using the \textit{nnAudio} \cite{cheuk_nnaudio_2020} library.
For training, we used the ADAM optimizer with a learning rate of $1$$\times$$10^{-4}$ and a batch size of $15$. For the most part, models are trained with an early stopping set at $200$ epochs. 
Depending on the training options, regularization terms are added with a multiplicative factor ($\alpha$, $\beta$, $\gamma$) so that each term has the same weight as \textit{$\mathcal{L}_{rec}$}.
The input to the mixture encoder is an STFT with a window size of $512$ and a hop size of $256$ samples, following \cite{schulze-forster_unsupervised_2023}.
HCQT representations used span $6$ octaves, with $60$ channels per octave, resulting in $20$ cents per frequency bin. The minimum frequency is set to $32.7$ Hz to ensure compatibility with the multi-F0 model in \cite{cuesta_multiple_2020}.  For the Multi-Scale Spectral loss ($\mathcal{L}_{rec}$), we used window sizes $\mathcal{N}=\{2048, 1024, 512, 256, 128, 64\}$.

\section{Datasets and experiments}
\label{Sec:Experiments}

\subsection{Datasets}
For our experiments, we use the different databases exploited in \cite{schulze-forster_unsupervised_2023}, namely: \textit{BC1Song}, \textit{BCBSQ} and \textit{Choral Singing Dataset}\footnote{All datasets are resampled at $16$ kHz; all audio mixes used are four seconds long allowing direct and consistent comparisons with the results of \cite{schulze-forster_unsupervised_2023}}.
A fourth database, \textit{Cantoría} is also used to test the generalization ability of our approaches. These four databases are briefly described below: 
\begin{itemize}[leftmargin=10px]
    \item \textit{Bach Chorales-Barbershop Quartet (BCBSQ)} is built from the databases Bach Chorales\cite{noauthor_pg_nodate-1} 
    and Barbershop Quartet \cite{noauthor_pg_nodate}. 
    It is a commercial database containing $26$+$22$ songs, performed by a quartet of singers (Soprano, Alto, Tenor and Bass (SATB) voices for the BC songs and tenor, lead, baritone and bass voices for BSQ). It contains recordings of all the individual voices, as well as recordings of the choir. This database is used for training with a total duration of $91$min $20$s and $9$min $10$s for validation. 

    \item \textit{BC1Song} is a reduced database that takes into account a single Bach Chorales song with a total duration of $2$min $40$s for training and $2$min $20$s for validation.

    \item \textit{Choral Singing Dataset (CSD)} \cite{cuesta_choral_2018} is a public database for choral singing. It consists of recordings of $3$ songs 
    performed by an SATB choir, with four singers per section ($4$ Soprano, $4$ Alto, $4$ Tenor and $4$ Bass). 
    The recordings of the 16 singers are separate, obtained from a spot mic and in a context where each section is recorded at the same time.
    Residues of the other $3$ singers in the section are then present in the different recordings. The database is only used for testing

    \item \textit{Cantoría} \cite{cuesta_cantoridataset_2022} is a choral database consisting of $11$ songs performed by a professional SATB choir. It contains recordings of all the voices as well as recordings containing the entire choir (Total duration of $36$min $10$s). 
    The dataset is only used for testing. 
\end{itemize}

\subsection{Experiments}


Our experiments aim at assessing the global performance of our model and how it compares to selected baselines under different training strategies. We only describe herein experiments where the multi-F0 estimation and voice assignment models are initialized with  pre-trained weights since a training of the complete architecture from scratch did not lead to satisfying results. We evaluate several training strategies as described below. When not specified otherwise, $\mathcal{L_{\text{full}}}$ is used for training (Eq. \ref{eq:full_loss}).

\begin{itemize}[leftmargin=10px]
    \item $S_FS_F$ where the models \textit{SalienceExtractor} and \textit{SalienceAssignment} are initialized with the pre-trained weights (from \cite{cuesta_multiple_2020} and\cite{cuesta_cantoridataset_2022}) and fixed during training (only $\mathcal{L_{\text{rec}}}$ is used);
    \item $S_{FT}S_{FT}$
    similar as $S_FS_F$ but after initialisation, the models are fine-tuned jointly with the training of the encoder/decoder; 
    

    \item $S_FS_{FT}$ where all submodules are pre-trained for initialisation and only the model \textit{SalienceExtractor} is fixed during training; 
    
    \item $W_{UP}$: in this setup, the assignment model is fine-tuned (with $\alpha \mathcal{L}_1 + \beta\mathcal{L}_2 + \gamma \mathcal{L}_3$) for $50$ epochs. Next, the set of models for F0 
    estimation is frozen, and the separation/synthesis model is trained for a further $50$ epochs ($\mathcal{L_{\text{rec}}}$ only). Finally, all modules are unfrozen and training continues using $\mathcal{L_{\text{full}}}$. 
\end{itemize}


We compare our results to the reference baseline UMSS \cite{schulze-forster_unsupervised_2023} and the U-Net model from \cite{petermann_deep_2020}.

\subsection{Evaluation}

The main metric used for evaluation is the Signal-to-Distortion Ratio (SI-SDR) \cite{roux_sdr_2019}, which measures the quality of the separated sources in relation to the original signals. 
\begin{equation}
    \text{SI-SDR} = 10\log_{10}\bigg(\frac{\Vert\eta s^2\Vert}{\Vert \eta s - \hat{s}^2 \Vert} \bigg),
    \label{eq:SI-SDR} 
\end{equation}
where $s$ denotes the target source, $\hat{s}$ the estimated source, and $\eta = \text{argmin}_{\eta} |\eta s - \hat{s}|^{2}$. 
The F0 estimation accuracy is evaluated using three metrics. The Raw Pitch Accuracy (RPA) measures the percentage of voiced frames in which the estimated pitch is within a $0.5$ semitone range of the ground-truth \cite{polinerMelodyTranscriptionMusic2007}. The Raw Chroma Accuracy (RCA) measures the same quantity as RPA but allows for octave errors. Finally, the Overall Accuracy (OA) takes into account all frames, including unvoiced cases. 


\section{Results}
\label{Sec:Results}
\subsection{Discussion}

Tables \ref{subtable_1} and \ref{subtable_2} show the experimental results for models trained on BC1Song and BCBSQ (larger database), respectively.
Out of the trained models, $S_FS_F$ has the worst performance. The addition of regularization in models $S_{FT}S_{FT}$, $S_FS_{FT}$, and $W_{UP}$ significantly improves performance across all metrics. 
Our best-performing model, $W_{UP}$, achieves $85$\% RPA (and $85$\% RCA) on BC1Song and $87$\% RPA  ($88$\% RCA) on BCBSQ. There is, however, a significant drop of approximately $8\%$ in the OA metric, indicating a tendency to incorrectly predict periodicity in unvoiced frames.

{
\setlength{\tabcolsep}{3pt} 
\small
 \begin{table}[]
 \centering
 \small
 \begin{subtable}[h]{\columnwidth}
    \centering
    \begin{tabular}{lllllllllll}
    
       \toprule
      Model &
      \multicolumn{2}{c}{SI\_SDR [dB] } &
      \multicolumn{2}{c}{OA [\%] } &
      \multicolumn{2}{c}{RPA [\%] } &
      \multicolumn{2}{c}{RCA [\%] }  \\
      \cmidrule(lr){2-3}
      \cmidrule(lr){4-5}
      \cmidrule(lr){6-7}
      \cmidrule(lr){8-9}
     &
      \multicolumn{1}{c}{$\mu$} &
      \multicolumn{1}{c}{Md} &
      \multicolumn{1}{c}{$\mu$} &
      \multicolumn{1}{c}{Md} &
      \multicolumn{1}{c}{$\mu$} &
      \multicolumn{1}{c}{Md} &
      \multicolumn{1}{c}{$\mu$} &
      \multicolumn{1}{c}{Md}  \\ \midrule
    UMSS \cite{schulze-forster_unsupervised_2023}           & \textbf{6.65} & \textbf{7.56} & -   & -  & -  & -  & -    & -    \\
    U-Net \cite{petermann_deep_2020}      & 1.5           & 2.72          & -  & -  & -  & -  & -  & -      \\ \midrule
    $S_FS_F$    & 2.93          & 3.59          & 66 & 68 & 72 & 75 & 73 & 77  \\
    $S_{FT}S_{FT}$ & 5.03          & 6.2           & 76 & 81 & 83 & 89 & 84 & 89  \\
    $S_FS_{FT}$ & 4.84          & 6.1           & 76 & 81 & 83 & 89 & 84 & 90  \\
    $W_{UP}$      & 5.22          & 6.34          & 77 & 82 & 85 & 90 & 85 & 91  \\ \bottomrule
    
    \end{tabular}%
    \caption{BC1Song \\ }
    
    \label{subtable_1}
    \end{subtable}

    \hfill
\vfill
\text{}
    \begin{subtable}[h]{\columnwidth}
    \centering
    \begin{tabular}{lllllllllll}
    \toprule
     Model &
      \multicolumn{2}{c}{SI\_SDR [dB] } &
      \multicolumn{2}{c}{OA [\%] } &
      \multicolumn{2}{c}{RPA [\%] } &
      \multicolumn{2}{c}{RCA [\%] }  \\
      \cmidrule(lr){2-3}
      \cmidrule(lr){4-5}
      \cmidrule(lr){6-7}
      \cmidrule(lr){8-9}
     &
      \multicolumn{1}{c}{$\mu$} &
      \multicolumn{1}{c}{Md} &
      \multicolumn{1}{c}{$\mu$} &
      \multicolumn{1}{c}{Md} &
      \multicolumn{1}{c}{$\mu$} &
      \multicolumn{1}{c}{Md} &
      \multicolumn{1}{c}{$\mu$} &
      \multicolumn{1}{c}{Md}  \\
     \midrule
    UMSS \cite{schulze-forster_unsupervised_2023}           & \textbf{6.91} & \textbf{7.60} & -  & -  & -  & -  & -  & -    \\
    U-Net \cite{petermann_deep_2020}      & 4.44           & 5.71          & -  & -  & -  & -  & -  & -      \\ \midrule
    $S_FS_F$    & 2.93          & 3.59           & 66 & 68 & 72 & 75 & 73 & 77  \\
    $S_{FT}S_{FT}$ & 4.81          & 6.07        & 73 & 79 & 80 & 87 & 82 & 88  \\
    $S_FS_{FT}$ & 5.77          & 6.46           & 78 & 82 & 85 & 90 & 85 & 89  \\
    $W_{UP}$      & 6.20         & 6.91          & 79 & 84 & 87 & 91 & 88 & 92  \\ \bottomrule
    \end{tabular}%
    \caption{BCBSQ}
    \label{subtable_2}
    \end{subtable}

    \caption{Evaluation of proposed approaches on  CSD (Test dataset) using 
    source separation and pitch accuracy 
    metrics, for the BC1Song (\subref{subtable_1}) and BCBSQ (\subref{subtable_2}) training datasets. $\mu$ stands for the mean and Md for the median. The models are trained and evaluated on mixtures with 4 sources. For all metrics, higher is better. 
    }
    \label{tab:résults-bc1song}
\end{table}
}

Our models consistently outperform Petermann et al.'s \cite{petermann_deep_2020} U-Net model in the SI-SDR metric. The best results were achieved by $W_{UP}$, scoring $5.52$ dB for BC1Song and $6.20$ dB for BCBSQ. However, they do not meet the performance benchmarks of our baseline \cite{schulze-forster_unsupervised_2023} ($6.65$ and $6.91$ dB), which uses pre-extracted F0 and manual source assignment. This trend holds across both datasets.
We believe that this slight performance decrease may be due to the differentiation strategy adopted for the voice assignment module with gradient copy.   
Models trained on the larger BCBSQ dataset have better performance, as shown in \autoref{subtable_2}. Unlike reported in \cite{schulze-forster_unsupervised_2023} we do not observe improved relative performance on smaller datasets. We hypothesize that larger datasets are helpful for F0 estimation generalization, as evidenced by the increase in RPA and RCA metrics on BCBSQ compared to BC1Song.

Integrating multi-F0 estimation during training adds complexity and impacts performance. Models like $W_{UP}$ and $S_FS_{FT}$, which use frozen salience extraction, yield better results, with $W_{UP}$ being the most effective. We observed that errors in F0 estimation have a cascading effect on the training and performance of the source separation model, which is consistent with findings in  \cite{schulze-forster_unsupervised_2023} and results in the seminal DDSP paper \cite{engel_ddsp_2020}, where even monophonic F0 joint estimation was reported to be a major challenge. 

\subsection{Ablation}

We discuss here in an ablation experiment the effectiveness of the proposed differentiable voice assignment module in the whole training process.  
This analysis is limited to the two best-performing approaches, $W_{UP}$ and $S_FS_{FT}$. For the warm-up approach ($W_{UP}$), we give the results obtained for each main stage of training: warm-up of the assignment model (\textit{F0}), training of the synthesis model (\textit{Synth}) and further training of the complete model which leads to ($W_{UP}$). For the $S_FS_{FT}$ approach, we continue training with the entire trainable architecture, which is referred to $S_FS_{FT}F$ in   \autoref{tab:Decomposition-phase-training}.

It can be seen that for both approaches, the training of the complete architecture is beneficial which confirms the effectiveness of our proposed contour extraction process. It also shows that the model used to extract the multi-frequency saliency maps (blocked for the $S_FS_{FT}$ method) becomes more efficient thanks to information derived from the Multi-Scale Spectral loss.

{\setlength{\tabcolsep}{2pt} 
\begin{table}[]
\centering
\begin{tabular}{lcccccccccc}
\toprule
  Model &
  \multicolumn{2}{c}{SI\_SDR [dB] } &
  \multicolumn{2}{c}{OA [\%] } &
  \multicolumn{2}{c}{RPA [\%] } &
  \multicolumn{2}{c}{RCA [\%] } \\
  \cmidrule(lr){2-3}
  \cmidrule(lr){4-5}
  \cmidrule(lr){6-7}
  \cmidrule(lr){8-9}
             & $\mu$         & Md            & $\mu$       & Md          & $\mu$       & Md          & $\mu$       & Md                  \\ \midrule
\textit{F0}           & 1.99          & 2.67          & 77          & 82          & 84          & 90          & 85          & 90         \\
\textit{Synth}        & 5.44          & 6.23          & 77          & 82          & 84          & 90          & 85          & 90                  \\
$W_{UP}$      & \textbf{6.2}  & \textbf{6.91} & \textbf{79} & \textbf{84} & \textbf{87} & \textbf{91} & \textbf{88} & \textbf{92} \\ \midrule
 $S_{FT}S_{FT}$     & 5.77          & 6.46          & 78          & 82          & 85          & 90          & 86          & 90          \\
$S_FS_{FT}F$ & \textbf{5.95} & \textbf{6.76} & \textbf{79} & \textbf{83} & \textbf{87} & \textbf{91} & \textbf{88} & \textbf{92}  \\ \bottomrule
\end{tabular}%
\caption{Evaluation of training stages for \textit{W-Up} and $S_FS_{FT}$ on BCBSQ  
}
\label{tab:Decomposition-phase-training}
\end{table}
}

\subsection{Generalization capabilities on a new dataset: Cantor\'ia}

We study herein the ability of our approach to generalize on another database (Cantor\'ia). The results presented in \autoref{tab:my-table} correspond to the performance on Cantor\'ia of the models trained on BC1song and BCBSQ. All scores are clearly lower, underlining the difficulty of the task, but our approach ($W_{UP}$) is clearly most robust in this context. We relate this to the generalisation capabilities of our voice assignment module. In the original model, the heuristical assignment model was very efficient, but apparently overfitted on the characteristics of the training databases BC1song and BCBSQ.

\begin{table}[h]
\centering
\begin{tabular}{lllll}
\toprule
Model & \multicolumn{2}{c}{BC1Song} & \multicolumn{2}{c}{BCBSQ} \\ \cmidrule(lr){2-3} \cmidrule(lr){4-5}
      & \multicolumn{1}{c}{$\mu$}  & \multicolumn{1}{c}{Md}    & \multicolumn{1}{c}{$\mu$}  & \multicolumn{1}{c}{Md} \\ \midrule
UMSS \cite{schulze-forster_unsupervised_2023} & 0.31  & 0.73 & 0.86  & 1.38 \\
U-Net \cite{petermann_deep_2020}              & -2.31 & -2.07 & 0.97  & 1.47 \\
$W_{UP}$                                         & \textbf{1.93} & \textbf{2.61} & \textbf{3.29} & \textbf{3.79} \\ \bottomrule
\end{tabular}
\caption{SI-SDR [dB] on Cantor\'ia for the models trained on BC1song and BCBSQ}
\label{tab:my-table}
\end{table}

To support reproducibility, we open source our code and provide a demo page with sound examples\footnote{Audio demo at: \href{https://pierrechouteau.github.io/umss\_icassp/audio}{https://pierrechouteau.github.io/umss\_icassp/audio} ; Code at \href{https://github.com/PierreChouteau/umss\_icassp}{https://github.com/PierreChouteau/umss\_icassp}}. 

\section{Conclusion}\label{sec:conclusion}

We have proposed a fully differentiable architecture for unsupervised music source separation which can be trained end-to-end. Our results demonstrate the merits of this new architecture and in particular they show that our model has better generalization capabilities when applied to more diverse data. Future work will be dedicated to the design of an architecture that could be efficiently trained from scratch, extending the current model which relies on a thoroughly designed warm-up procedure with pre-trained sub-modules.   

\balance
\vfill\eject
\bibliographystyle{IEEEtran}
\bibliography{refs, icaasp_differentiable}

\end{document}